\documentclass[aip,jap,superscriptaddress,reprint]{revtex4-1}

\usepackage{graphicx}
\usepackage{epstopdf} 
\usepackage{amssymb} 
\usepackage{amsmath} 

\renewcommand\vec[1]{{\bf #1}}

\begin{document}

\title{Remote Surface Roughness Scattering in FDSOI devices with high-$\kappa$/SiO$_2$ gate stacks}

\author{Y. M. Niquet}
\email{yniquet@cea.fr}
\affiliation{CEA, INAC-SP2M, L\_Sim, 17 rue des Martyrs, 38054 Grenoble, France}
\affiliation{Univ. Grenoble Alpes, 17 rue des Martyrs, 38054 Grenoble, France}
\author{I. Duchemin}
\affiliation{CEA, INAC-SP2M, L\_Sim, 17 rue des Martyrs, 38054 Grenoble, France}
\affiliation{Univ. Grenoble Alpes, 17 rue des Martyrs, 38054 Grenoble, France}
\author{V.-H. Nguyen}
\affiliation{CEA, INAC-SP2M, L\_Sim, 17 rue des Martyrs, 38054 Grenoble, France}
\affiliation{Univ. Grenoble Alpes, 17 rue des Martyrs, 38054 Grenoble, France}
\author{F. Triozon}
\affiliation{CEA, LETI, MINATEC Campus, 17 rue des Martyrs, 38054 Grenoble, France}
\affiliation{Univ. Grenoble Alpes, 17 rue des Martyrs, 38054 Grenoble, France}
\author{D. Rideau}
\affiliation{STMicroelectronics, 850 Rue Jean Monnet, 38920 Crolles, France}


\begin{abstract}
We investigate remote surface scattering (RSR) by the SiO$_2$/HfO$_2$ interface in Fully-Depleted Silicon-on-Insulator (FDSOI) devices using Non-Equilibrium Green's Functions. We show that the RSR mobility is controlled by cross-correlations between the surface roughness profiles at the Si/SiO$_2$ and SiO$_2$/HfO$_2$ interfaces. Therefore, surface roughness and remote surface roughness can not be modeled as two independent mechanisms. RSR tends to enhance the total mobility when the Si/SiO$_2$ interface and SiO$_2$ thickness profiles are correlated, and to decrease the total mobility when they are anti-correlated. We discuss the implications for the high-$\kappa$/Metal gate technologies.
\end{abstract}

\maketitle

High-$\kappa$ materials such as HfO$_2$ have been introduced in Metal-Oxide-Semiconductor Field Effect Transistors in order to keep a tight electrostatic control over short channels while limiting gate leakage currents.\cite{Robertson06} They usually remain separated from the channel by a thin interfacial layer (IL) of SiO$_2$. Yet the introduction of high-$\kappa$ materials leads to a systematic decrease of carrier mobilities, which is very significant at weak inversion densities, but can persist in the strong inversion regime.\cite{Casse06} 

Different scattering mechanisms have been put forward to explain this degradation. There is no doubt that charges trapped at the SiO$_2$/HfO$_2$ interface or in the HfO$_2$ layer (Remote Coulomb scattering (RCS)) make a major contribution at weak inversion.\cite{Gamiz03, Esseni03b, Toniutti12} Remote scattering by polar optical phonons (RPH) in HfO$_2$ is also a serious candidate, especially in thin IL devices.\cite{Fischetti01, Toniutti12} Much less attention has been given up to now to scattering by roughness at the SiO$_2$/HfO$_2$ interface or equivalently by IL thickness fluctuations. At variance with RCS, this mechanism, known as remote surface roughness (RSR) scattering, is expected to be dominant in the strong inversion regime.  

RSR has first been investigated with semi-classical Kubo-Greenwood approaches in a different context, namely roughness at the SiO$_2$/Gate interface in polysilicon gate technologies.\cite{Jia87, Walczak01, Gamiz03b, Saito04} The models were later extended to HfO$_2$ and HfO$_2$/IL gate stacks.\cite{Ghosh06, Zhang07} However, the different interfaces were assumed uncorrelated, and surface roughness (SR)/remote surface roughness modeled as two independent mechanisms.

In this letter, we use Non-Equilibrium Green's Functions (NEGF) methods\cite{Niquet14, Nguyen14} to investigate RSR in the latest Fully-Depleted Silicon-on-Insulator (FDSOI) thin-film technologies with high-$\kappa$/Metal gates. We show that RSR scattering is dominated by roughness at the SiO$_2$/HfO$_2$ interface, indeed prevails at large carrier densities (at variance with RCS), and decreases exponentially with IL thickness. More importantly, we demonstrate that the RSR correction is controlled by cross-correlations between the Si/SiO$_2$ and SiO$_2$/HfO$_2$ interfaces. SR and RSR can not, therefore, be modeled as two independent mechanisms. RSR is beneficial when the Si/SiO$_2$ interface and the IL thickness profiles are correlated, and detrimental if they are in anti-correlated.

Let us first discuss the basics of remote surface scattering. The main effect of RSR is to modulate oxide thicknesses hence the effective field within the semiconductor. This gives rise to potential and charge fluctuations responsible for extra carrier scattering.\cite{Jia87, Walczak01, Gamiz03b, Saito04, Ghosh06, Zhang07} In this picture, we expect stronger scattering from the SiO$_2$/HfO$_2$ interface than from the HfO$_2$/Metal gate interface, since the latter is usually much farther from the channel and is efficiently screened by the HfO$_2$ layer. This will indeed be confirmed by our numerical simulations on FDSOI devices. From now on, we hence focus on RSR at the SiO$_2$/HfO$_2$ interface.

We can draw the main trends and identify the relevant structural parameters from semi-classical perturbation theory. Let $\Delta(\vec{r})$ and $\Delta^\prime(\vec{r})$ be the SR profiles at the Si/SiO$_2$ and SiO$_2$/HfO$_2$ interfaces, and $\delta(\vec{r})=\Delta^\prime(\vec{r})-\Delta(\vec{r})$ be the variations of the thickness of the IL. According to Fermi Golden Rule, the scattering rate between states $|\vec{k}\rangle$ and $|\vec{k}^\prime\rangle$ with in-plane wave vectors $\vec{k}$ and $\vec{k}^\prime$ is proportional to $\overline{\left|\langle\vec{k}^\prime|H_{\rm R}|\vec{k}\rangle\right|^2}$, where $H_{\rm R}$ is the Hamiltonian of the SR+RSR disorder, the over-line denotes an average over configurations, and we have discarded band indexes for simplicity. This matrix element can be expanded in powers of $\Delta$ and $\delta$:
\begin{equation}
\langle\vec{k}^\prime|H_{\rm R}|\vec{k}\rangle=A(\vec{q})\Delta(\vec{q})+B(\vec{q})\delta(\vec{q})+...\,,
\end{equation}
where $\vec{q}=\vec{k}^\prime-\vec{k}$, $\Delta(\vec{q})$ and $\delta(\vec{q})$ are the Fourier transforms of $\Delta(\vec{r})$ and $\delta(\vec{r})$, and $A(\vec{q})$ and $B(\vec{q})$ are complex numbers. Hence, to second order in $\Delta$ and $\delta$,
\begin{multline}
\overline{\left|\langle\vec{k}^\prime|H_{\rm R}|\vec{k}\rangle\right|^2}=|A(\vec{q})|^2\overline{\left|\Delta(\vec{q})\right|^2}+|B(\vec{q})|^2\overline{\left|\delta(\vec{q})\right|^2}\\
+A^*(\vec{q})B(\vec{q})\overline{\Delta^*(\vec{q})\delta(\vec{q})}+{\rm c.c.}
\label{eqRSR}
\end{multline}
The first term will be referred to as ``pure SR'' scattering (SR scattering is, indeed, usually computed\cite{Goodnick85, Jin09} with assumptions best compatible\cite{noteRSR} with $\delta=0$). It is proportional to $\overline{\left|\Delta(\vec{q})\right|^2}$, which is the Fourier transform of the auto-covariance function $F_\Delta(\vec{R})=\overline{\Delta(\vec{r})\Delta(\vec{r}+\vec{R})}$ of the SR profile at the Si/SiO$_2$ interface ($\vec{r}$ and $\vec{R}$ being in-plane vectors). The second term is ``uncorrelated'' RSR scattering by fluctuations of the IL thickness. It is, likewise, proportional to the Fourier transform of the auto-covariance function $F_\delta(\vec{R})=\overline{\delta(\vec{r})\delta(\vec{r}+\vec{R})}$ of these fluctuations.\cite{Jia87, Walczak01, Gamiz03b, Saito04, Ghosh06, Zhang07} The third (and fourth complex conjugate) terms are ``cross-correlated'' contributions to RSR scattering. They are proportional to $\overline{\Delta^*(\vec{q})\delta(\vec{q})}$, which is the Fourier transform of the correlation function $G(\vec{R})=\overline{\Delta(\vec{r})\delta(\vec{r}+\vec{R})}$ between the SR profile at the Si/SiO$_2$ interface and the IL thickness fluctuations. They have not been accounted for in Refs. \onlinecite{Jia87, Walczak01, Gamiz03b, Saito04, Ghosh06, Zhang07}.

In general, one expects $|B(\vec{q})|\ll|A(\vec{q})|$, so that the cross-correlations prevail over the uncorrelated term unless $G(\vec{R})\sim 0$ -- which is unlikely in the thin ILs where RSR might be significant. The above analysis therefore shows that SR and RSR can not be modeled as two independent mechanisms, as assumed in previous literature.

We consider an exponential (auto-)correlation model for the Si/SiO$_2$ interface profile $\Delta(\vec{r})$ and IL thickness variations $\delta(\vec{r})$:\cite{Goodnick85}
\begin{eqnarray}
&F_\Delta(\vec{R})=\Delta^2 f(R)\ ;\ F_\delta(\vec{R})=\delta^2 f(R)& \nonumber \\
&G(\vec{R})=\Delta\delta f(R) g\,,&
\label{eqF}
\end{eqnarray}
where $f(R)=e^{-\sqrt{2}R/L_c}$ and $L_c=1.3$ nm everywhere. $g$ characterizes the cross-correlations (or similarity) between $\Delta(\vec{r})$ and $\delta(\vec{r})$ ($|g|\le 1$). This is equivalent to the following model for the SiO$_2$/HfO$_2$ interface profile $\Delta^\prime(\vec{r})$:
\begin{eqnarray}
F_{\Delta^\prime}(\vec{R})&=&\overline{\Delta^\prime(\vec{r})\Delta^\prime(\vec{r}+\vec{R})}=\Delta^{\prime 2} f(R) \nonumber \\
H(\vec{R})&=&\overline{\Delta(\vec{r})\Delta^\prime(\vec{r}+\vec{R})}=\Delta\Delta^\prime f(R) h\,,
\end{eqnarray}
where $\Delta^{\prime 2}=\Delta^2+\delta^2+2\Delta\delta g$ and $h=(\Delta+\delta g)/\Delta^\prime$ is the overlap between the two interfaces. They are parallel if $h=1$, uncorrelated if $h=0$, and ``anti-parallel'' if $h=-1$ [Conversely, $\delta^2=\Delta^2+\Delta^{\prime 2}-2\Delta\Delta^\prime h$ and $g=(\Delta^\prime h-\Delta)/\delta$]. Although the parameters $\Delta^\prime$ and $h$ are more intuitive,\cite{Palasantzas03} the equivalent ($\delta$, $g$) set is more practical. Indeed, pure SR scattering simply corresponds to $\delta=0$; Also, this set provides a straightforward criterion for the integrity of the SiO$_2$ layer: The surface covered by holes ($\delta(\vec{r})<-t_{\rm IL}$) in an IL with nominal thickness $t_{\rm IL}$ is $\lesssim 2.3\%$ as long as $\delta<t_{\rm IL}/2$.\cite{noteIL} The present results and conclusions might not, therefore, hold if $\delta\gg t_{\rm IL}/2$.

The squared matrix element Eq. (\ref{eqRSR}) thus reads:
\begin{multline}
\overline{\left|\langle\vec{k}^\prime|H_{\rm R}|\vec{k}\rangle\right|^2}=f(\vec{q})\left\{|A(\vec{q})|^2\Delta^2 \right. \\
+\left.\vphantom{|A(\vec{q})|^2}2\Re e\left[A(\vec{q})B^*(\vec{q})\right]\Delta\delta g\right\}\,,
\label{eqRSR2}
\end{multline}
where we have neglected the $\propto |B(\vec{q})|^2\delta^2$ term.

Finally, let $\mu_{\rm SR}$ be the SR-only mobility (computed for strictly parallel interfaces,\cite{noteRSR} i.e. $\delta=0$) and $\mu_{\rm SR+RSR}$ the total SR+RSR mobility. Let us define the RSR mobility as $\mu_{\rm RSR}^{-1}=\mu_{\rm SR+RSR}^{-1}-\mu_{\rm SR}^{-1}$. Then, we expect from Eq. (\ref{eqRSR2}) in a semi-classical Kubo-Greenwood approach:
\begin{eqnarray}
\mu_{\rm RSR}^{-1}\simeq -C_{\rm RSR}^{-1}\Delta\delta g\,,
\label{eqmuRSR}
\end{eqnarray}
where $C_{\rm RSR}$ is a constant. The RSR mobility is therefore inversely proportional to the rms fluctuations $\Delta$ and $\delta$, and to the overlap $g$. Note that $\mu_{\rm RSR}$ (as defined above) can be positive or negative, i.e. remote surface roughness can either enhance or inhibit the flow of carriers depending on $g$. We will discuss the underlying physics later.

In order to verify these assertions, we have performed Non-Equilibrium Green's Functions (NEGF) calculations of the mobility in $n$FDSOI devices along the lines of Ref. \onlinecite{Niquet14} (same structures). We have considered $t_{\rm Si}=4$ and $t_{\rm Si}=8$ nm thick (001) Si films (dielectric constant $\varepsilon_{\rm Si}=11.7$) on a 25 nm thick buried oxide and a grounded $n$-doped Si substrate. The front gate stack is made of an IL of SiO$_2$ with varying thickness $t_{\rm IL}$ ($\varepsilon_{\rm IL}=3.9$), a 2.4 nm thick layer of HfO$_2$ ($\varepsilon_{\rm HfO_2}=22$), and a metal gate. We use the effective mass approximation for electrons. The wave functions can penetrate in the IL. Electron-phonon (PH) interactions are included in all calculations, but cancel out in $\mu_{\rm RSR}$ (see Ref. \onlinecite{Niquet14} for the methodology and a discussion about Matthiessen's rule).


\begin{figure}
\includegraphics[width=\columnwidth]{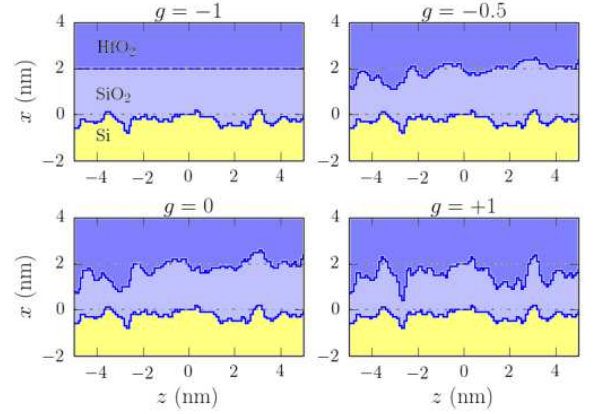} 
\caption{Surface roughness profiles computed for different $g$'s ($\Delta=\delta=0.47$ nm and $t_{\rm IL}=2$ nm). Note that the SiO$_2$/HfO$_2$ interface is actually smooth when $\Delta=\delta$ and $g=-1$. The $g=-0.5$ stack approximately lies on the curve ``$\Delta^\prime=0.47$ nm, $\lambda=2.6$ nm'' of Fig. \ref{Figmutotal}.}     
\label{FigDens}
\end{figure}

The advantage of NEGF is that it makes use of an explicit, real space description of SR/RSR disorders and screening (no need for an approximation to the very complex $H_{\rm R}$ as in Kubo-Greenwood calculations). This allows for accurate calculations of SR mobilities.\cite{Niquet14} In the present case, we have computed mobilities in samples of SR/RSR disorder satisfying Eqs. (\ref{eqF}) (see Fig. \ref{FigDens}), and averaged over a few configurations until convergence.

\begin{figure}
\includegraphics[width=\columnwidth]{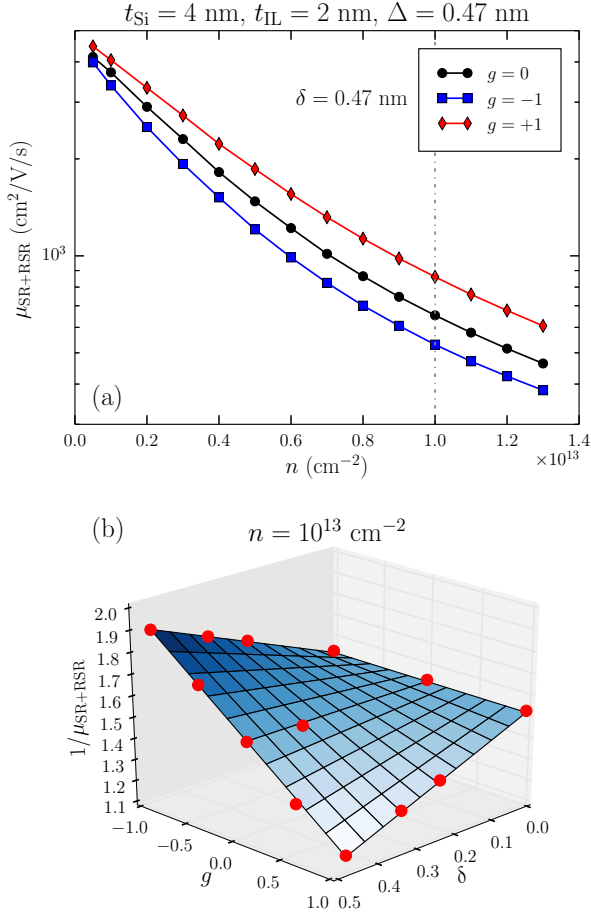} 
\caption{(a) $\mu_{\rm SR+RSR}$ as a function of carrier density $n$ and $g$ in a $n$FDSOI device with $t_{\rm Si}=4$ nm and $t_{\rm IL}=2$ nm ($\Delta=\delta=0.47$ nm) (b) $\mu_{\rm SR+RSR}^{-1}$ ($10^{-3}$ V.s/cm$^2$) as a function of $\delta$ (nm) and $g$ in the same device, at $n=10^{13}$ cm$^{-2}$. The red dots are the NEGF data, the blue surface is the best fit $\mu_{\rm SR+RSR}^{-1}\equiv\mu_{\rm SR}^{-1}+C_{\rm RSR}^{-1}\Delta\delta g$.}     
\label{FigRSRdg}
\end{figure}

As an illustration, the total SR+RSR mobility $\mu_{\rm SR+RSR}^{-1}\equiv\mu_{\rm SR}^{-1}+\mu_{\rm RSR}^{-1}$ is plotted as a function of carrier density $n$ and $g$ in Fig. \ref{FigRSRdg}a, and as a function of $\delta$ and $g$ in Fig. \ref{FigRSRdg}b. As expected from Eq. (\ref{eqmuRSR}), $\mu_{\rm SR+RSR}^{-1}$ is pretty linear with $\delta$ and $g$, and does not show, in particular, very significant $\propto\delta^2$ contributions from the uncorrelated term in Eq. (\ref{eqRSR}). We have also checked that RSR at the HfO$_2$/Metal gate interface had much less impact on the mobility, at least for the HfO$_2$ thickness $\ge 2$ nm considered here.

\begin{figure}
\includegraphics[width=\columnwidth]{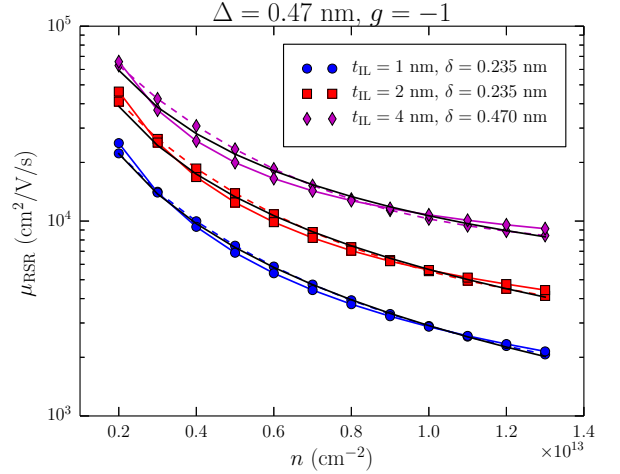} 
\caption{RSR mobilities in 4 nm thick (dashed color lines with symbols) and 8 nm thick (solid color lines with symbols) Si films with $\Delta=0.47$ nm and $g=-1$. Black lines are Eqs. (\ref{eqmuRSR}, \ref{eqCRSR}).}     
\label{FigmuRSR}
\end{figure}

The RSR mobility computed for $g=-1$ is plotted as a function of carrier density for different film and IL thicknesses in Fig. \ref{FigmuRSR}. Since $\mu_{\rm RSR}>0$ when $g<0$, the total mobility decreases when the fluctuations of the IL thickness are anti-correlated to the SR profile. Indeed, the carriers, squeezed by the surface electric field $F_s$, tend to localize where $\Delta(\vec{r})>0$; localization and scattering are enhanced by the concomitant increase of $F_s$ when $g<0$ (IL tends to thin where $\Delta(\vec{r})>0$), and are, conversely, inhibited by the concomitant decrease of $F_s$ when $g>0$ (IL tends to thicken where $\Delta(\vec{r})>0$). As expected, the RSR mobility decreases with increasing carrier density (or equivalently $F_s$) as the carriers move closer to the top gate stack; Hence RSR is most efficient (as is surface roughness) in the strong inversion regime, at variance with RCS for example. As such, it is weakly dependent on the thickness of the silicon film in the investigated range $t_{\rm Si}\ge 4$ nm. It shows, though, a strong (exponential) dependence on the thickness of the IL; the thinner the IL the stronger the RSR correction.

From these calculations, we can extract the prefactor $C_{\rm RSR}(n, t_{\rm IL})$ as a function of the carrier density $n$ and thickness of the IL. The Fourier components of the electric field suggest\cite{Jia87, Zhang07} that $C_{\rm RSR}^{-1}\sim Ke^{-\alpha t_{\rm IL}}$. Expanding $K$ and $\alpha$ in powers of $n$, we tentatively fit the NEGF data with
\begin{equation}
C_{\rm RSR}^{-1}=an\left(1+\frac{n}{n_1}\right)\exp\left[-\left(1+\frac{n}{n_2}\right)\frac{t_{\rm IL}}{t_0}\right]\,,
\label{eqCRSR}
\end{equation}
and get $a=2.96\times 10^{-2}$ V.s/cm$^2$, $n_1=9.6\times 10^{12}$ cm$^{-2}$, $t_0=1.87$ nm, and $n_2=4.1\times 10^{13}$ cm$^{-2}$. Eq. (\ref{eqCRSR}) reproduces the NEGF data very well for $n>2\times 10^{12}$ cm$^{-2}$, as shown on Fig. \ref{FigmuRSR}.

\begin{figure}
\includegraphics[width=\columnwidth]{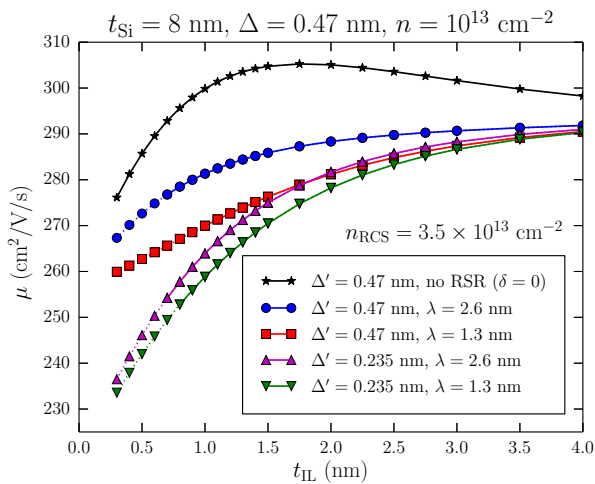} 
\caption{Total mobility (PH+SR+RSR+RCS) in a 8 nm thick $n$FDSOI film as a function of $t_{\rm IL}$, for different rms roughness $\Delta^\prime$ at the SiO$_2$/HfO$_2$ interface, and for different vertical correlation lengths $\lambda$ [Eq. (\ref{eqh})] ($\Delta=0.47$ nm, $n_{\rm RCS}=3.5\times 10^{13}$ cm$^{-2}$). The carrier density is $n=10^{13}$ cm$^{-2}$ (effective field $E_{\rm eff}\sim 0.8$ MV/cm). The data were interpolated from a set of NEGF calculations. The lines are dotted wherever $\delta>t_{\rm IL}/2$, i.e. when the IL might be pierced in many places.}
\label{Figmutotal}
\end{figure}

How large can RSR scattering be in actual devices ? As an illustration, we make the following assumption for the correlations between the Si/SiO$_2$ and SiO$_2$/HfO$_2$ interfaces:\cite{Palasantzas03}
\begin{equation}
h(t_{\rm IL})=e^{-t_{\rm IL}/\lambda}\,,
\label{eqh}
\end{equation}
where $\lambda$ is a vertical correlation length. The total PH+SR+RSR+RCS mobility is plotted a function of $t_{\rm IL}$ in Fig. \ref{Figmutotal}, for a 8 nm thick film in the strong inversion regime, and for different $\lambda$ and $\Delta^\prime$. RCS was modeled as a distribution of positive charges at the SiO$_2$/HfO$_2$ interface\cite{Niquet14, Nguyen14} with density $n_{\rm RCS}=3.5\times 10^{13}$ cm$^{-2}$. In the absence of RSR, the mobility is weakly dependent on the IL thickness, at variance with experimental data.\cite{Ragnarsson11b} The RSR can, however, significantly enhance scattering in thin ILs. It is, notably, very detrimental if the SiO$_2$/HfO$_2$ interface is smoother than the Si/SiO$_2$ interface, because the IL thickness fluctuations become anti-correlated to the Si/SiO$_2$ roughness profile ($\delta\to\Delta$ and $g\to -1$ when $\Delta^\prime\to 0$). Note that alloying SiO$_2$ with HfO$_2$ near the IL/High-$\kappa$ interface might blur the variations of the dielectric constant and effectively smooth that interface. A more detailed comparison with experiment\cite{Ragnarsson11b} would call for a comprehensive de-embedding of the different scattering mechanisms.\cite{Nguyen14} Yet we can conclude that RSR, together with RPH\cite{Fischetti01, Toniutti12} (not accounted for at present in NEGF), are serious candidates in explaining the trends measured in very thin IL devices at strong inversion. 

Finally, we would like to discuss the effects of non-parabolic (NP) corrections and exchange-correlation (XC) effects. NP corrections can be included using a two bands $\vec{k}\cdot\vec{p}$ model\cite{Sverdlov08, Niquet14} for the conduction band, yet at a much larger numerical cost. Long-range XC effects have been modeled by an image charge self-energy correction (numerically computed along the lines of Ref. \onlinecite{Bescond10}), and short-range XC effects by a local density approximation.\cite{Jin09, Hedin71} Calculations on a few representative configurations show that NP and XC corrections both decrease the mobility and $C_{\rm RSR}$ (i.e., enhance RSR scattering), by a total of $\sim20\%$.


To conclude, we have discussed the impact of remote surface roughness scattering (RSR) at the SiO$_2$/HfO$_2$ interface on the electron mobility in FDSOI thin films. We show that the RSR corrections are dominated by the cross-correlations between the two interfaces of the IL. The mobility increases when the SiO$_2$ thickness fluctuations are correlated with those of the Si/SiO$_2$ interface, and decreases otherwise. SR and RSR can not, therefore, be described as two independent mechanisms, and must be modeled concurrently. The RSR mobility shows an exponential dependence on the IL thickness. In particular, RSR might be significant at strong inversion in thin IL devices where the SiO$_2$/HfO$_2$ interface is (dielectrically) smoother than the Si/SiO$_2$ interface.

This work was supported by the French National Research Agency (Project ``Noodles''). The NEGF calculations were run on the TGCC/Curie machine using allocations from PRACE and GENCI.

\providecommand{\noopsort}[1]{}\providecommand{\singleletter}[1]{#1}%


\begin{thebibliography}{23}
\bibitem [{not({\natexlab{a}})}]{noteRSR}%
  \BibitemOpen
  \bibinfo {note} {In line with the
  standard assumptions made in SR mobility calculations,\cite{Goodnick85,Jin09} $\mu_{\rm
  SR}\equiv\mu(\delta=0)$ is free from scattering by oxide thickness fluctuations -- the main
  driving force for RSR. It does, however, include ``Prange-Nee''-like
  scattering\cite{Jin09} by the SiO$_2$/HfO$_2$ interface (weak except in the
  thinnest ILs).}\BibitemShut {Stop}%
\bibitem [{not({\natexlab{b}})}]{noteIL}%
  \BibitemOpen
  \bibinfo {note} {This condition results
  from the fact that $\delta(\vec{r})$, for given $\vec{r}$, is a normal random
  variable.}\BibitemShut {Stop}%
\end{thebibliography}
\end{document}